\documentclass[prd,aps,a4paper,preprint,preprintnumbers,nofootinbib]{revtex4-1}
\usepackage[a4paper, hdivide={1.91cm,,1.165cm}, vdivide={1.83cm,,3.3cm}]{geometry}

\usepackage{amsmath,amssymb}
\usepackage{graphicx,multirow}
\usepackage{tabularx}
\usepackage[export]{adjustbox}
\usepackage[hyperfootnotes=false]{hyperref}
\usepackage[usenames, dvipsnames]{color}
\usepackage[normalem]{ulem}
\usepackage{float}

\newcommand{\ii}{\ensuremath{\text{i}}}
\newcommand{\dd}{\ensuremath{\text{d}}}

\newcommand{\beq}{\begin{eqnarray}}% can be used as {equation} or  {eqnarray}
\newcommand{\eeq}{\end{eqnarray}}

\begin{document}
%%%%%%%%%%%%%%%%%%%%%%%%%%%%%%%%%

\preprint{UCI-TR-2021-14}

\title{The Ubiquity of Gauged Q-Shells}

\author{Julian~Heeck}
\email{heeck@virginia.edu}
\affiliation{Department of Physics, University of Virginia,
Charlottesville, Virginia 22904-4714, USA}

\author{Arvind~Rajaraman}
\email{arajaram@uci.edu}
\affiliation{Department of Physics and Astronomy, 
University of California, Irvine, CA 92697-4575, USA
}

\author{Christopher~B.~Verhaaren}
\email{cverhaar@uci.edu}
\affiliation{Department of Physics and Astronomy, 
University of California, Irvine, CA 92697-4575, USA
}

\begin{abstract}
Non-topological gauged soliton solutions called Q-balls arise in many scalar field theories that are invariant under a $U(1)$ gauge symmetry. The related, but qualitatively distinct, Q-shell solitons have only been shown to exist for special potentials. We investigate gauged solitons in a generic sixth-order polynomial potential (that contains the leading effects of many effective field theories) and show that this potential generically allows for both Q-balls and Q-shells. We argue that Q-shell solutions occur in many, and perhaps all, potentials that have  previously only been shown to contain Q-balls. We  give simple analytic characterizations of these Q-shell  solutions, leading to excellent predictions of their physical properties.
\end{abstract}

%%%%%%%%%%%%%%%%%%%%%%%%%%%%%%%%%
\maketitle
\newpage
\tableofcontents
%%%%%%%%%%%

\section{Introduction\label{s.intro}}

Non-topological solitons~\cite{Lee:1991ax,Nugaev:2019vru}, herein simply referred to as solitons, correspond to classical field configurations carrying a conserved Noether charge $Q$. {Stable} solitons furthermore have a smaller energy than $Q$ individual charges and therefore cannot disperse, although decays into smaller solitons might be possible.

The simplest examples of solitons arise in $U(1)$-invariant field theories of complex scalars, which can form Q-balls~\cite{Coleman:1985ki,Heeck:2020bau}.
Promoting the $U(1)$ symmetry to a gauge symmetry complicates the differential equations but still allows for Q-ball soliton solutions over part of the parameter space~\cite{Lee:1988ag,Gulamov:2013cra,Gulamov:2015fya,Heeck:2021zvk}. Even for simple scalar potentials, the underlying field equations are impossible to solve analytically and have to be approached either numerically or using analytic approximations, as in Ref.~\cite{Heeck:2021zvk}, for example.

Because of the difficulty in solving the coupled differential equations for gauged solitons, most studies have focused on solutions that are qualitatively similar to the global Q-ball case, so similar, in fact, that there exists a mapping between the global and gauged solitons~\cite{Heeck:2021zvk}. However, these need not be the only possible gauged soliton solutions:
Ref.~\cite{Arodz:2008nm} identifies gauged soliton solutions where the radial scalar profile does not form a ball but rather a thin shell; these Q-\emph{shell} solitons were studied for the somewhat unrealistic but simple V-shaped potential~\cite{Arodz:2008nm}, log potential~\cite{Tamaki:2014oha,Panin:2016ooo}, and, recently, in multi-field models~\cite{Ishihara:2021iag}.
Compared to the rather mature field of Q-ball studies, Q-shells have been explored far less, in part due to the difficulty in finding them numerically and in part because it is unclear which potentials admit Q-shell solutions.

Here, we argue that, contrary to the implications of these special potentials, Q-shells arise in many, if not all, potentials that produce Q-balls. We demonstrate how a simple sixth-order polynomial potential leads to both gauged Q-balls and gauged Q-shells. Despite many studies of this potential over the years, to our knowledge, these Q-shell solutions have never been discussed in the literature. We find that the qualities of this potential that give rise to Q-shells are found in a vast number of potentials. Therefore, on general grounds, we expect Q-shell solutions to be more generic than previously anticipated and conjecture that they can arise in most scalar potentials that allow for global Q-balls. In addition to numerical solutions, we provide analytic approximations that describe these new solitons to very good accuracy and allow us to fully explore the gauged-soliton parameter space of this potential.

In Sec.~\ref{s.gaugedQ}, we review the basics of non-topological gauged solitons, and the equations that describe them. Section~\ref{s.theory} outlines how the various solitons, Q-ball and Q-shell, can be understood as particle trajectories. In doing so, we make use of the language of a particle rolling in a two-dimensional potential. We derive an approximate analytical Q-shell solution is Sec.~\ref{s.qshells}, from which quantities like the size, energy, and charge of the Q-shells can be predicted. We also use these approximate solutions to determine, in Sec.~\ref{s.cond}, what conditions must be satisfied to support Q-shell solitons. In Sec.~\ref{s.results}, we compare our predictions to the exact numerical results, finding excellent agreement, before concluding in Sec.~\ref{s.con}.

\section{Gauged Solitons\label{s.gaugedQ}}

We study gauged solitons that result from the Lagrangian density
\begin{equation}
\mathcal{L}=\left|D_\mu\phi \right|^2-U(|\phi|)-\frac14 F_{\mu\nu}F^{\mu\nu},
\end{equation}
where $D_\mu=\partial_\mu-\ii e A_\mu$ is the gauge covariant derivative and $F_{\mu\nu}=\partial_\mu A_\nu
-\partial_\nu A_\mu$ the field-strength tensor. The parameter $e$ is the gauge coupling, normalized so that the complex scalar $\phi$ has unit charge.
The $U(1)$-symmetric potential is only required to have the property that $U(|\phi|)/|\phi|^2$ has a minimum at $|\phi|=\phi_0/\sqrt{2}>0$ such that 
\begin{equation}
0\leq \frac{\sqrt{2U(\phi_0/\sqrt{2})}}{\phi_0}\equiv\omega_0<m_\phi\,,\label{e.Omega0}
\end{equation} 
where $m_\phi$ is the mass of $\phi$. The location of the minimum, $\phi_0$, is chosen positive without loss of generality.
We furthermore demand $\langle \phi\rangle =0$ in vacuum and choose the potential energy to be zero in vacuum.
Following the notation and conventions of Ref.~\cite{Heeck:2021zvk}, we make the static charge ansatz~\cite{Lee:1988ag} 
\begin{equation}
\phi (t,\vec{x})=\frac{\phi_0}{\sqrt{2}} f(r) e^{\ii\,\omega t}\,,\qquad A_0(t,\vec{x})=\phi_0 A(r)\,,\qquad A_i(t,\vec{x})=0\,,\label{e.fdef2}
\end{equation}
and define the dimensionless quantities 
\begin{align}
\rho &\equiv r\sqrt{m_\phi^2-\omega_0^2}\,, &
\Phi_0 &\equiv\frac{\phi_0}{\sqrt{m_\phi^2-\omega_0^2}}\,, &
 \alpha &\equiv e\Phi_0\,,\\
\Omega &\equiv\frac{\omega}{\sqrt{m_\phi^2-\omega_0^2}}\,, &
\Omega_0 &\equiv\frac{\omega_0}{\sqrt{m_\phi^2-\omega_0^2}}\,, &
\kappa^2 &\equiv \Omega^2 - \Omega_0^2\,.
\label{eq:rescaling}
\end{align}
The Lagrangian can then be rewritten in terms of the dimensionless functions $f$ and $A$ as
\begin{equation}
L=4\pi\Phi_0^2\sqrt{m^2_\phi-\omega_0^2}\int \dd\rho\,\rho^2\left\{ -\frac12f^{\prime 2}+\frac{1}{2}A^{\prime2}
+\frac{1}{2}f^2\left(\Omega-\alpha A \right)^2-\frac{U(f)}{\Phi_0^2(m_\phi^2-\omega_0^2)^2}\right\}.\label{e.massesLag}
\end{equation}
The scalar frequency $\omega$ is restricted to the region $\omega_0<\omega \leq m_\phi$, which translates into the allowed range $0< \kappa \leq 1$.
The Lagrangian $L$ for the two scalar fields $f$ and $A$ can be interpreted as a single particle moving under the influence of the two-dimensional potential
 \beq
 V(f,A)=\frac{1}{2}f^2\left(\Omega-\alpha A \right)^2-\frac{U(f)}{\Phi_0^2(m_\phi^2-\omega_0^2)^2}\,,
 \label{eq:potentialV}
 \eeq
 with $\rho$ playing the role of the time coordinate.
In this analogy, however, the $A$ field's kinetic term has 
the wrong sign, meaning that the particle rolls uphill in the $A$ direction. As an example, note that for a fixed $f>0$, the potential in $A$ has a minimum at
\beq
A_m=\frac{\Omega}{\alpha}~.\label{e.Amax}
\eeq
The sign of the $A$ kinetic term implies that if $A\geq A_m$ it either feels a `force' pushing it to larger and larger values of $A$ or no force in the $A$ direction at all. Because we are interested in solitons whose gauge field falls off to zero as $\rho\to\infty$, this proves that $A(\rho)< A_m$ for all soliton solutions. 

To fully define and obtain localized soliton solutions, the equations of motion that result from the Lagrangian,
\begin{align}
f'' + \frac{2}{\rho} f'&= -\frac{\partial V}{\partial f} = \frac{1}{\Phi_0^2(m_\phi^2-\omega_0^2)^2}\frac{\dd U}
{\dd f}-\left(\Omega-\alpha A\right)^2f\,,\label{e.feq}\\
A'' + \frac{2}{\rho} A' &= +\frac{\partial V}{\partial A} = \alpha f^2( \alpha A-\Omega)\,,\label{e.Aeq}
\end{align}
are solved subject to the boundary conditions $f'(0)=A'(0)=0$ and
\begin{equation}
 \lim_{\rho\to\infty}f = \lim_{\rho\to\infty}A =0 \,.
\end{equation}
The right-hand side of the $A$ equation illustrates, in agreement with the potential analogy, that the gauge field decreases monotonically for $A(\rho)<A_m$ and $f\neq 0$. 

The soliton's conserved charge $Q$ is obtained by integrating the time component of the scalar 
current over all space~\cite{Lee:1988ag}:
\begin{align}
Q&=4\pi \Phi_0^2\int \dd\rho\,\rho^2f^2\left(\Omega-\alpha A \right) . \label{e.charge}
\end{align}
Because $A(\rho)< A_m$, we see that $Q$ is always positive, in agreement with our normalizing the $\phi$ charge to one. We can use the $A$ equation \eqref{e.Aeq} to rewrite the soliton charge as
\beq
Q=-\frac{4\pi\Phi_0^2}{\alpha}\lim_{\rho\to\infty}\rho^2A'~.
\eeq
This agrees with the simple expectation that outside the soliton the leading behavior of the gauge field is to fall off like $Q/\rho$. The energy $E$ of the soliton is given by the Hamiltonian~\cite{Lee:1988ag,Heeck:2021zvk}
\begin{align}
E/\sqrt{m_\phi^2-\omega_0^2} & =4\pi\Phi_0^2\int \dd\rho\,\rho^2\left\{ \frac12f^{\prime 2}+\frac{1}{2}A^{\prime2}
+\frac{1}{2}f^2\left(\Omega-\alpha A \right)^2+\frac{U(f)}{\Phi_0^2(m_\phi^2-\omega_0^2)^2}\right\}\label{e.Eint}\\
&= \Omega Q+\frac{4\pi\Phi_0^2}{3}\int \dd\rho\,\rho^2\left(f^{\prime 2}-A^{\prime2} \right) \label{e.energy}.
\end{align}

In most of what follows, we confine our discussion to the most generic $U(1)$-symmetric sextic potential, conveniently parametrized as
\begin{align}
U(f) = \phi_0^2\frac{f^2}{2} \left[ (m_\phi^2 -\omega_0^2)(1-f^2)^2 +\omega_0^2 \right]=\Phi_0^2\left(m_\phi^2-\omega_0^2\right)^2\frac{f^2}{2} \left[(1-f^2)^2 +\Omega_0^2 \right] .
\label{eq:sextic_potential}
\end{align}

%%%%%%%%%%%%%%%%%%%%%%%%%%%%%%%
\section{Gauged Soliton Trajectories\label{s.theory}}

Solitons in the potential~\eqref{eq:sextic_potential}
have been thoroughly discussed for global~\cite{Heeck:2020bau} and gauged~\cite{Lee:1991ax,Heeck:2021zvk} Q-balls. 
We show here that this theory admits a different class of gauged solitons\textemdash Q-\emph{shells}. These are solutions
where the charge density close to the origin vanishes or nearly vanishes.
While Q-shells have been found in other nongeneric
potentials~\cite{Arodz:2008nm,Tamaki:2014oha,Panin:2016ooo}, to our knowledge, it has not previously been shown that the general 
potential~\eqref{eq:sextic_potential} admits Q-shells.

The solutions to the differential equations~\eqref{e.feq} and~\eqref{e.Aeq} can be understood qualitatively by studying the effective two-dimensional potential of Eq.~\eqref{eq:potentialV}. For our sextic potential~\eqref{eq:sextic_potential}, it takes the form
\begin{align}
V(f,A)=\frac{1}{2}f^2\left[\kappa^2-\alpha A(2\Omega-\alpha A)-\left(1-f^2\right)^2 \right] .
\label{eq:eff_potential}
\end{align}
This potential is a function of $f^2$, so we can focus our attention on the $f\geq0$ region. For each value of $A$, this region has three extrema in $f$, two at
\beq
f^2_\pm=\frac13\left(2\pm\sqrt{1+3\kappa^2-3\alpha A(2\Omega-\alpha A)} \right) ,\label{e.fpm}
\eeq
with $f_-$ a minimum and $f_+$ a maximum, and one maximum at $f=0$. 
Gauged solitons can then be related to a particle rolling in this two-dimensional potential. 

\begin{figure}[t]
\includegraphics[width=0.49\textwidth]{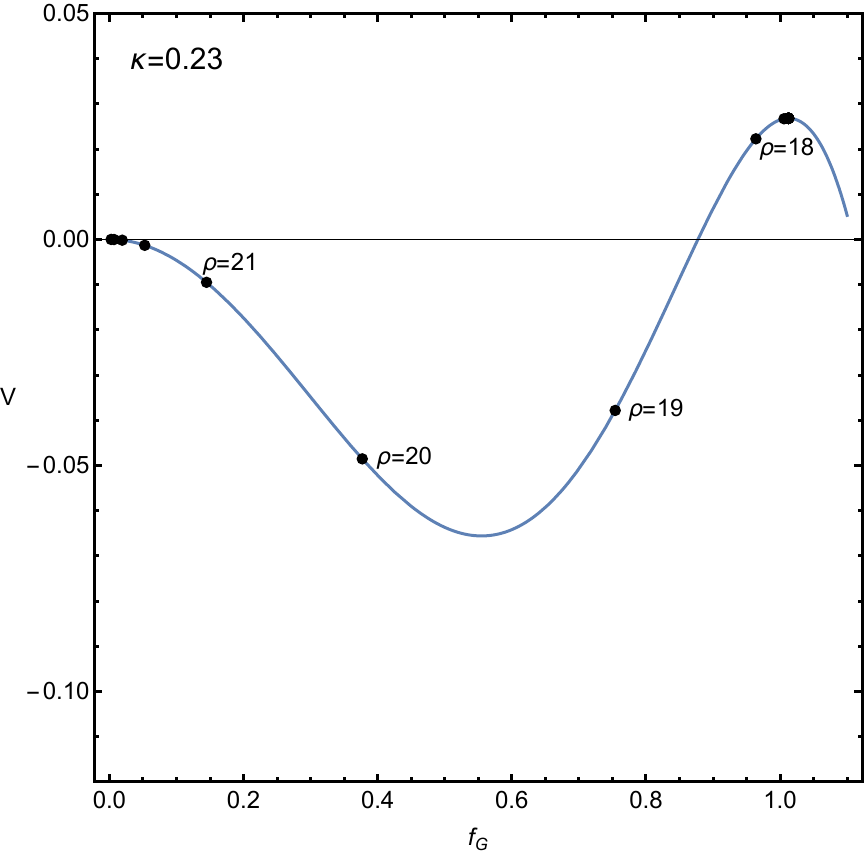}
\includegraphics[width=0.49\textwidth]{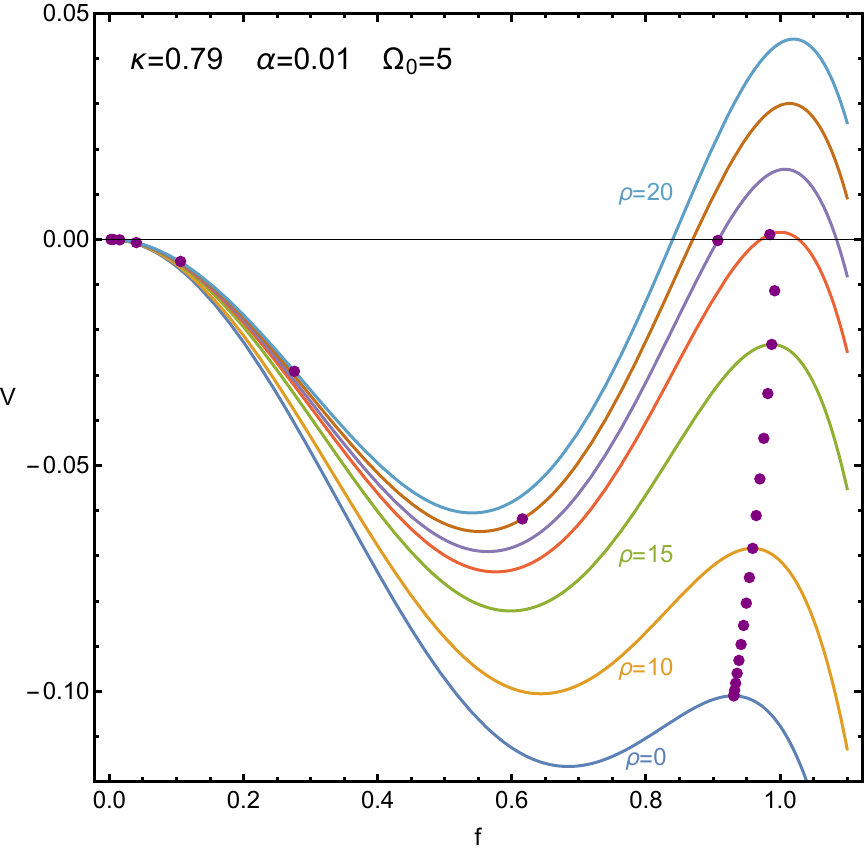}
\caption{Black (purple) points indicate the values of the global (gauged) Q-ball profiles for integer values of $\rho\in[0,25]$. 
\emph{Left:}
Effective potential for the global Q-ball
 \emph{Right:} Effective potentials for gauged Q-ball $f$ at specific values of $A(\rho)$.  
}
\label{fig:gaugepotentialPlot}
\end{figure}

For global Q-balls ($\alpha=0$) in the thin-wall regime, the particle
starts at the maximum at $f_+$ and then rolls quickly down to the maximum at $f=0$; i.e.~the scalar field has a constant value $f_+$ out to a large radius, with a quick transition to $f=0$. An example of a global potential is shown on the left side of Fig.~\ref{fig:gaugepotentialPlot}; the black dots show the value of $f$ for discrete values of $\rho$. The function begins near $f_+\gtrsim1$ until $\rho\sim18$ and then quickly rolls through the valley and back up near $f=0$. While we have used the sextic potential for concreteness, the essential characteristics of Q-ball solutions are quite general. The particle begins at or near a local maximum, the location and shape of which are determined by the particular potential, and then rolls to the local maximum at $f=0$. Thick-wall Q-balls do not start near the maximum but at some point, with $V>0$, down the potential hill. Consequently, nothing impedes their immediate rolling, with friction, through the valley to the maximum at $f=0$.

\begin{figure}[t]
\includegraphics[width=0.49\textwidth]{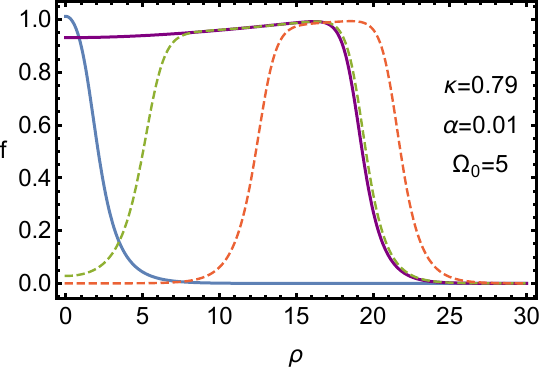}
\includegraphics[width=0.49\textwidth]{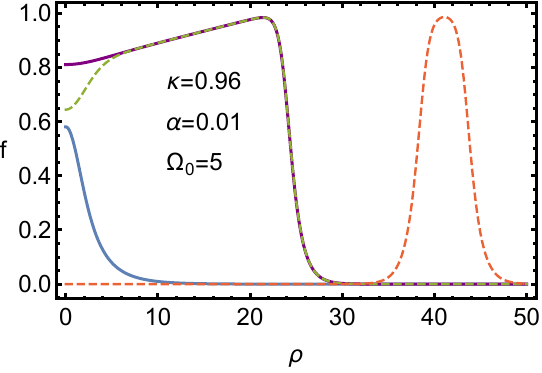}\\
\includegraphics[width=0.49\textwidth]{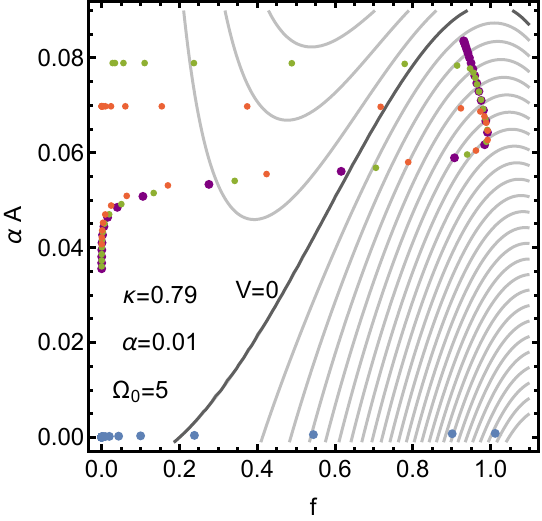}
\includegraphics[width=0.49\textwidth]{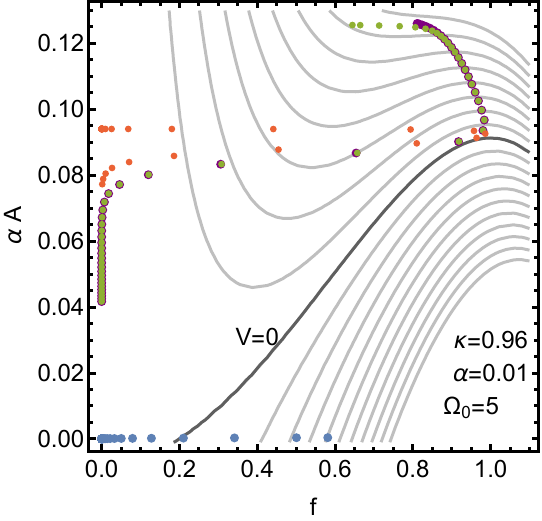}
\caption{
Examples of the four soliton profiles (above), Q-balls (solid) and Q-shells (dashed), that can exist for a given set of potential parameters, and the trajectory of these profiles on the potential (below). The potential plots show contours of constant $V$, with the $V=0$ contour shown in black. The trajectory points for the same profiles (coded by color) are integer values of $\rho$, from $\rho=0$ to $\rho=30$ (left) or $\rho=50$ (right).
}
\label{fig:profiles}
\end{figure}

In the case of gauged Q-balls, the effective potential 
 changes while the particle rolls because the gauge field evolves as a function of $\rho$. As shown in the previous section, $A$ decreases monotonically from some value $A(0)<\Omega/\alpha$ for localized solitons. For large enough $A$, the local maximum at $f_+$ can be below the local maximum at
$f=0$. As $A$ evolves, the value of the potential at the $f_+$ maximum increases until it rises near to the value of the $f=0$ maximum. Once this occurs, the particle can roll from near $f=f_+$ to $f=0$, similar to the global case. An example of such a profile for a gauged Q-ball is shown in the right panel of Fig.~\ref{fig:gaugepotentialPlot}, with the purple points denoting values of $f$ for discrete $\rho$. Several effective potentials for a given $A(\rho)$ are also shown. We see that the particle remains near $f_+$ even as the value of $V(f_+)$ increases. When $\rho\sim 17$, the particle begins to roll along the potential until it reaches $f=0$. The $f$ profile for this trajectory is shown in purple within the top left panel of Fig.~\ref{fig:profiles}. The full trajectory in $f$ and $A$ through a contour plot of $V$ is shown in purple within the lower left panel of the same figure.

In contrast to global Q-balls, there can be more than one gauged soliton for the same parameter values. 
As is well known, for a given set or parameters, there are generally between zero and two Q-balls solutions~\cite{Heeck:2021zvk}. These are given by the purple and blue curves in Fig.~\ref{fig:profiles}. Typically, one of these has a large radius (purple curve), and one has a small radius (blue curve);
the large radius Q-ball is often referred to as a
thin-wall Q-ball (the duration of the transition from $f_+$ to zero is much smaller that the Q-ball size), and
the small radius Q-ball is referred to as  a
thick-wall Q-ball. Both Q-balls are analyzed in detail in Ref.~\cite{Heeck:2021zvk} and can be understood as being mapped from the global Q-ball solitons for a given potential. Notice that the thick-wall Q-ball begins with small $A(\rho)$, so the potential maximum near $f_+$ is already greater than the $f=0$ maximum. This is why the particle simply rolls down the hill, very much like the global case, with the gauge field providing no appreciable effect. As the trajectories on the right-side of Fig.~\ref{fig:profiles} show, the particle can begin some way down the potential hill, already away from the maximum at $f_+$.

Beyond these two Q-ball solutions, an entirely different class of gauged solitons exists; the
particle can start at or near the local maximum at $f=0$ and roll down to the
maximum at $f_+$. This can only occur if the $f_+$ maximum is
below the one at $f=0$. The evolution of the gauge field, see Eq.~\eqref{e.Aeq}, is suppressed when $f$ is small, meaning that before the particle rolls from the $f=0$ maximum the gauge field remains nearly constant. Once the particle rolls down, the gauge field begins to decrease, and the maximum at $f_+$
can rise above the maximum at $f=0$, similar to the thin-wall Q-balls. The scalar field can then roll back to the minimum at
$f=0$. Such particle trajectories correspond to a Q-shell: the scalar field is near zero at the origin,
goes to $f_+$ at an intermediate radius, and then goes back to zero.
The existence of the local maximum at $f_+$ is not crucial for this kind of solution, and the same discussion can be made for effective potentials that only feature the local maximum at $f=0$ and a minimum at $f>0$~\cite{Tamaki:2014oha}.
Since these are the basic properties the potential needs to fulfill to admit even global soliton solutions, we expect gauged Q-shells to exist in many if not all potentials that lead to 
Q-{balls}. Restrictions on the potential parameters are derived in Sec.~\ref{s.cond} for the sextic potential given in Eq.~\eqref{eq:sextic_potential}.

\begin{figure}[t]
\includegraphics[width=0.49\textwidth]{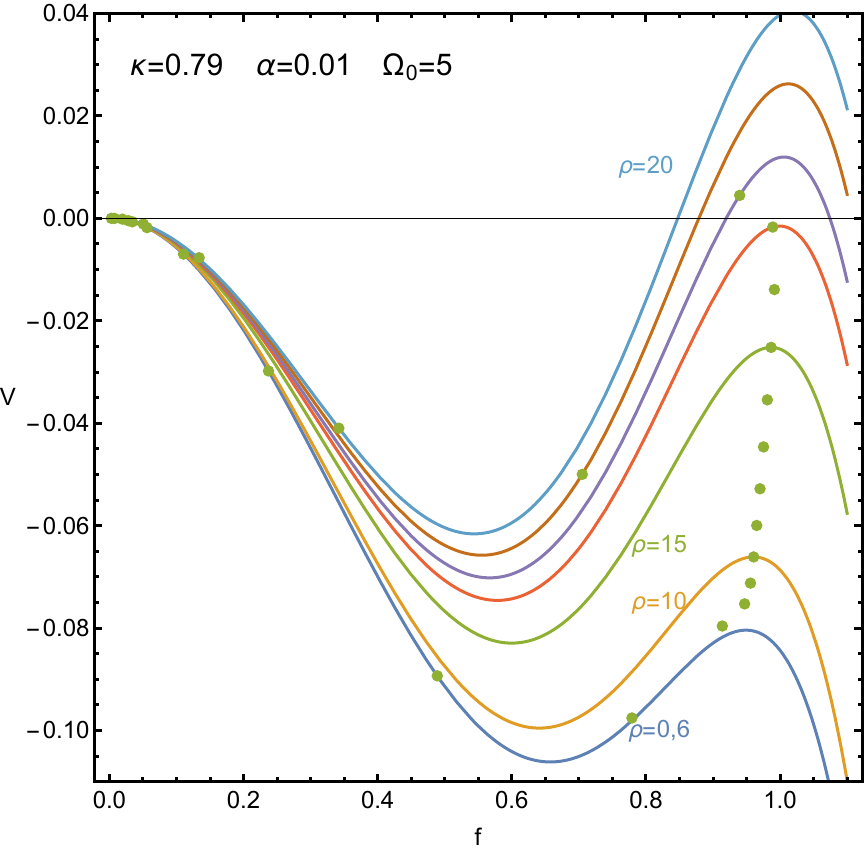}
\includegraphics[width=0.49\textwidth]{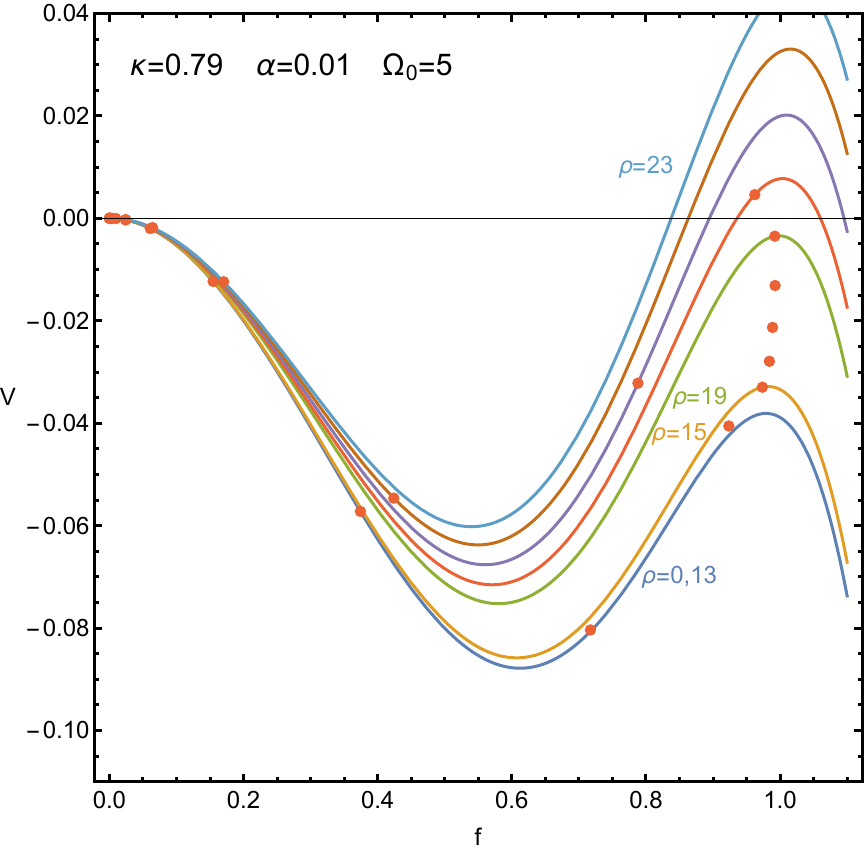}
\caption{Green (orange) points indicate the values of the wide (narrow) gauged Q-shell profiles for integer values of $\rho\in[0,25]$. The colored curves show the effective potentials for $f$ at specific values of $A(\rho)$.
\emph{Left:}
The wide Q-shell trajectory begins near $f=0$ and then rolls down the hill toward the local maximum ($f_+$) nearer to $f=1$. It remains near $f_+$ until $V>0$ and then rolls back toward $f=0$.
 \emph{Right:} The narrow Q-shell trajectory is similar to the wide but remains near $f=0$ until larger $\rho$ and spends a shorter duration in $\rho$ at $f_+$.  
}
\label{fig:gaugeShellPotentialPlot}
\end{figure}

For this potential, as we derive in the following section, there are up to two distinct Q-shell solitons for a given parameter point. For instance, as shown on the left panel of Fig.~\ref{fig:gaugeShellPotentialPlot},
$f$ can start near the maximum at $f=0$ and quickly (in $\rho$) roll down to the maximum $f_+$ where it follows the thin Q-ball solution shown in the right panel of Fig~\ref{fig:gaugepotentialPlot}.
This leads to {\it wide} Q-shells, which are shown in green in Fig.~\ref{fig:profiles}, which also shows how closely it follows the thin-wall Q-balls shown in purple.
Alternatively,
$f$ can remain near $f=0$ for a long time (that is, until large $\rho$) before rolling up to $f_+$ for a short time and then rolling back to zero, as shown in the right panel of Fig.~\ref{fig:gaugeShellPotentialPlot}.
This leads to {\it narrow} Q-shells which in Fig.~\ref{fig:profiles} are shown in orange. 

Figure~\ref{fig:profiles} also shows that the final transition toward $f=0$ for both Q-shells occurs near the same place in the effective potential as the thin-wall Q-ball transition. This is shown for two values of $\kappa$, with the larger (on the right) producing a narrow Q-shell that is completely separate in $\rho$ from the other solitons. Although the transitions occur at different $\rho$, and hence with different amounts of friction, they nevertheless begin near the same point in the potential. This can be understood from energy arguments outlined in the following section. The right panel of the figure also illustrates that not all solitons begin at a maximum in the effective potential. Both the thick-wall Q-ball (blue) and the wide Q-shell (green) begin significantly downhill from the $f=f_+$ and $f=0$ maxima, respectively. 

Having found that up to four distinct gauged soliton solutions can exist for the same potential parameters, one may wonder if there are still more. This seems unlikely, as every soliton discovered so far is associated with the particle rolling from one maximum to another. In the case of Q-balls, it is from $f_+$ to $f=0$, and for Q-shells, it is from $f=0$ to $f_+$ and back to $f=0$. What about a further cycle, that is, $f_+$ to near $f=0$ and back to $f_+$ and then finally up to $f=0$? The problem is that $A$ decreases monotonically whenever $f$ is not small, and this makes the $f_+$ maximum grow. This means rolling from $f_+$ to near $f=0$ and back will put the particle farther from $f_+$ and unable to be lifted up by the growing maximum,  hence never having enough energy to reach $f=0$. This seems to forbid solitons with more than two transitions. However, this argument may fail in more complicated potentials, where additional maxima may play a role, see Refs.~\cite{Tamaki:2014oha,Panin:2016ooo} for instance.  We leave a detailed analysis of this possibility to future work.

\section{Q-Shell Profiles\label{s.qshells}}

In this section, we make the characterization of gauged Q-shells more precise. The preceding section indicates that Q-shells are related to thin-wall Q-balls. This motivates
 describing Q-shells, to first approximation, by a generalization of
the thin-wall Q-ball ansatz
\beq
f(\rho)=\left\{\begin{array}{lc}
0 & \rho<R_< \\
1 & \;\; R_<\leq\rho\leq R_> \\
0 & R_><\rho\,
\end{array} \right. ~,
\eeq 
where $R_<$ and $R_>$ correspond to the inner and outer radii of the Q-shell, respectively. 
Q-balls correspond to the special case $R_< =0$, whereas Q-shells are characterized by $R_< > 0$.
Using this ansatz, we solve the equation of motion for $A$ and find
\beq
A(\rho) =\left\{\begin{array}{cc}
A_<  & \rho< R_<\\
\displaystyle\frac{\Omega}{\alpha}-\frac{A_1}{\rho}e^{\alpha\rho} -\frac{A_2}{\rho}e^{-\alpha\rho} & R_< \leq \rho \leq R_>\\
\displaystyle A_>\frac{R_>}{\rho} & R_><\rho
\end{array} \right. ,
\label{e.Afield}
\eeq
for constants $A_<,\,A_>\,, A_1,$ and $A_2$. These are specified by demanding the gauge profile and its first derivative be continuous. However, to fully specify the profile, we also need to estimate the radii $R_<$ and $R_>$. 

For Q-balls, in which the inner region is absent, the transition region $ \rho\sim R_>$ is well described
by the transition profile~\cite{Heeck:2020bau},
\beq
f_T=\frac{1}{\sqrt{1+2e^{2(\rho-R_>)}}}~,
\eeq
where $\rho=R_>$ is the radius of the Q-ball, defined by $f''(R_>)=0$.
For Q-shells, we have two transitions, and the scalar profile can be approximated
as the product of two transition functions,
\begin{align}
f(\rho)=\frac{1}{\sqrt{1+2e^{2(R_<-\rho)}}}\frac{1}{\sqrt{1+2e^{2(\rho-R_>)}}} \,.
\label{e.ffield}
\end{align}
This functional form is remarkably successful in describing Q-shells and approximates the numerical solutions (see, for example, Fig.~\ref{fig:profiles}) very well.
The transition profiles also imply a relation between
 the change in effective-potential energy and the work done by friction during the transition~\cite{Heeck:2020bau}. This yields
 the work\textendash energy relation from Refs.~\cite{Heeck:2020bau,Heeck:2021zvk},
\begin{align}
V(f(\rho_<),A(\rho_<))-V(f(\rho_>),A(\rho_>)) +\frac{f'(\rho_<)^2-f'(\rho_>)^2}{2}-\frac{A'(\rho_<)^2-A'(\rho_>)^2}{2}  =2\int\displaylimits_{\rho_<}^{\rho_>}\frac{\dd\rho}{\rho}\left(f^{\prime2}-A^{\prime2} \right) ,
\end{align}
that holds for all gauged solitons.
Around the two transitions (i.e.~$\rho_< \lesssim R_< \lesssim \rho_>$ and $\rho_< \lesssim R_> \lesssim \rho_>$ with $f'(\rho_{>,<}) =0$),  $A$ is nearly constant, and we recover results reminiscent of the global Q-ball case,
\beq
R_<=-\frac{f^2_+(A_<)}{2V(f_+(A_<),A_<)}~, \ \ \ \ R_>=\frac{f^2_+(A_>)}{2V(f_+(A_>),A_>)}~,
\eeq
where the difference in sign comes from the first transition rolling away from the origin, while the second transition rolls toward it. Using the definition of $f_+$ in Eq.~\eqref{e.fpm} and the effective potential in Eq.~\eqref{eq:eff_potential}, we can rewrite  the radius relations as
\beq
R_<=\frac{1}{2f^2_+(A_<)[1-f^2_+(A_<)]}\,, \ \ \ \ R_>=-\frac{1}{2f^2_+(A_>)[1-f^2_+(A_>)]} \,.
\label{e.fullRshells}
\eeq
These equations only apply when $A_{<}$ is sufficiently small to keep $f_+$ real ($A_><A_<$ because $A$ is monotonically decreasing). At the largest possible $A$, we find that $f^2_+$ takes on its minimal real value of $2/3$, which implies that 
\beq
R_<>\frac{9}{4}
\label{e.kappaMax}
\eeq
for all allowed Q-shell solutions. 
The two equations in~\eqref{e.fullRshells} provide the remaining two relations among the radii and the parameters in the Lagrangian but are difficult to solve in complete generality. 

Similar to the global case~\cite{Heeck:2020bau}, the radii equations are approximated quite well by
\beq
R_<\simeq -\frac{1}{\kappa^2-\alpha A_<(2\Omega-\alpha A_<)}\,, \ \ \ \ R_>\simeq\frac{1}{\kappa^2-\alpha A_>(2\Omega-\alpha A_>)}\,,\label{e.Rshells}
\eeq
in the limit of large radii, which are easier to solve.
Along with the
four equations coming from
 continuity of $A$ and $A'$ at $R_<$ and $R_>$,
these relations determine the six unknown parameters: $A_1,\,A_2,\,A_<,\,A_>,\,R_<,\,$ and $R_>$. 
The first relation of Eq.~\eqref{e.Rshells}, along with continuity of $A$ and $A'$ at $R_<$, leads to
\begin{align}
A_< &= \frac{\Omega}{\alpha}-\frac{1}{\alpha}\sqrt{\Omega_0^2-\frac{1}{R_<}}~,\\
A_1 &= e^{-\alpha R_<}\frac{\alpha R_<+1}{2\alpha^2}\sqrt{\Omega_0^2-\frac{1}{R_<}}~,\\
A_2 &= e^{\alpha R_<}\frac{\alpha R_<-1}{2\alpha^2}\sqrt{\Omega_0^2-\frac{1}{R_<}}~\label{e.A2inside}
\end{align}
while the continuity of $A$ and $A'$ at $R_>$ and the second relation in Eq.~\eqref{e.Rshells} leads to
\begin{align}
A_> &= \frac{\Omega}{\alpha}-\frac{1}{\alpha}\sqrt{\frac{1}{R_>}+\Omega_0^2}~,\\
A_1 &= e^{-\alpha R_>} \frac{\alpha R_>\sqrt{\frac{1}{R_>}+\Omega_0^2}+\Omega}{2\alpha^2}~,\\
A_2 &= e^{\alpha R_>}\frac{\alpha R_>\sqrt{\frac{1}{R_>}+\Omega_0^2}-\Omega}{2\alpha^2}~.
\end{align}
Of course, the constants $A_1$ and $A_2$ must be the same, which leads to the two equations that determine the Q-shell radii,
\begin{align}
2 &= \left[\alpha_0^2\Sigma   (1+ \Delta )-\kappa_0^2\right]( \Sigma  - \Delta ) \,, 
\label{e.const1}\\
e^{-\alpha_0 \Delta  } &=\sqrt{1-\frac{2 }{ \Sigma  -\Delta }}\frac{1+\frac{\alpha_0}{2} ( \Sigma -\Delta  )}{\sqrt{1 + \kappa_0^2 }+\frac{\alpha_0}{2}(\Sigma  + \Delta )\sqrt{1+\frac{2}{\Sigma +\Delta }}}~,\label{e.const2}
\end{align}
where we have defined
\beq
\Delta\equiv \Omega_0^2(R_>-R_<)~, \ \ \ \ \Sigma\equiv \Omega_0^2(R_>+R_<)~, \ \ \ \ \alpha_0\equiv\frac{\alpha}{\Omega_0^2}~, \ \ \ \ \kappa_0\equiv\frac{\kappa}{\Omega_0}~.
\eeq
The implicit invariance of the radii equations under an appropriate rescaling of the parameters by powers of $\Omega_0$ is not a symmetry of the Lagrangian or even Eq.~\eqref{e.fullRshells} but is a result of the large-radii approximation in Eq.~\eqref{e.Rshells}. As such, this reduction of parameters is not expected to hold over the entire parameter space but clearly simplifies the analysis significantly.

Equation~\eqref{e.const1} is a quadratic equation and can be used to
find up to two solutions for $ \Sigma $ as a function of $ \Delta $. Then, Eq.~\eqref{e.const2} becomes a transcendental equation for $ \Delta $ as a function of $\alpha_0$ and $\kappa_0$. 
While an analytic solution of this equation is impossible, numerical solutions are easy to obtain and show that up to two different branches of $\Delta$ exist.

\begin{figure}[t]
\includegraphics[width=0.49\textwidth]{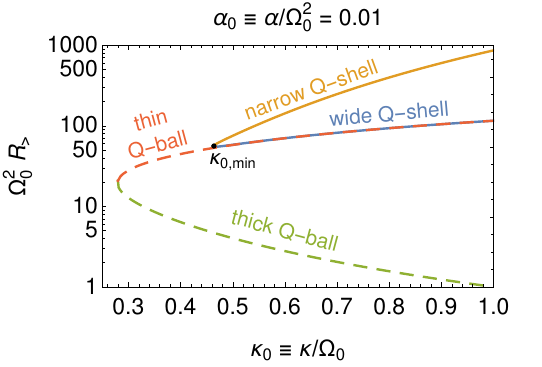}
\includegraphics[width=0.49\textwidth]{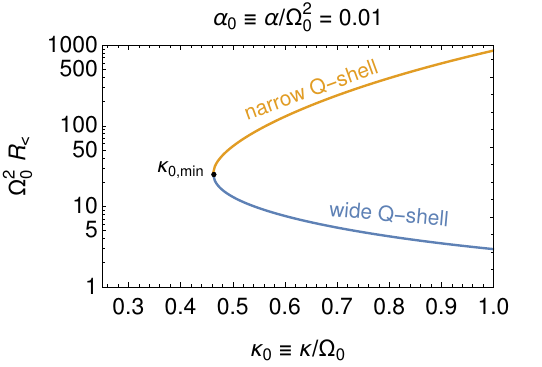}
\caption{
Predicted values of the outer (left) and inner (right) gauged soliton radii $R_{>,<}$ as a function of $\kappa_0$ for $\alpha_0=0.01$. 
Narrow Q-shells are shown in yellow, and wide Q-shells in blue.
Thin and thick Q-balls have $R_< = 0$ and are shown in dashed red and green, respectively.
The Q-shell branches merge at $\kappa_{0,\text{min}}$.
The wide Q-shell solutions for $R_>$ are degenerate with the thin Q-balls for $\kappa_0\gg \kappa_{0,\text{min}}$; both have a maximal $\kappa_0$ determined by Eq.~\eqref{e.kappaMax}, not indicated in the figure because it depends on $\Omega_0$.. 
}
\label{fig:Radii}
\end{figure}

In total, we have up to two sets of radii $0 < R_< < R_>$ that solve Eqs.~\eqref{e.const1} and~\eqref{e.const2} and correspond to Q-\emph{shells}, illustrated in Fig.~\ref{fig:Radii}. For Q-\emph{balls}, we fix $R_< =0$ and only solve the $R_>$ equation in Eq.~\eqref{e.Rshells} to find up to two solutions for $R_>$. A better Q-ball prediction can be obtained via the mapping of Ref.~\cite{Heeck:2021zvk}, but the current approach is sufficient for large radii.
Fig.~\ref{fig:Radii} shows our predictions for the Q-ball and Q-shell radii according to the procedure above. Notable features include the merging of narrow and wide Q-shells at a point $\kappa_{0,\text{min}}$, below which Q-shells cease to exist. Furthermore, the outer radius (as well as energy and charge) of wide Q-shells is degenerate with thin Q-balls for $\kappa_0 \gg \kappa_{0,\text{min}}$; even around $\kappa_0 \sim \kappa_{0,\text{min}}$, the deviation between these two branches is very small. This matches the expectation from the potential analysis above.

\section{Conditions for Q-Shells\label{s.cond}}

With the analytical approximations for profiles and radii in hand, we can discuss the required conditions for the potential parameters to admit Q-shells.
First, Eq.~\eqref{e.A2inside} shows that $\Omega_0 =0$ is not allowed for $R_< > 0$. Q-shells thus require $\Omega_0 >0$, which we assume in what follows. Demanding Eq.~\eqref{e.A2inside} to be real yields the lower bound $R_< > 1/\Omega_0^2$ for Q-shells.

For \emph{narrow} Q-shells, we have $\Delta \ll \Sigma$, so we can expand Eqs.~\eqref{e.const1} and~\eqref{e.const2} in small $\Delta$; expanding furthermore in small $\alpha_0$, we find the simple solutions
\begin{align}
\Delta \simeq \frac{2}{\kappa_0^2}\left( 1+ \sqrt{1+\kappa_0^2}\right) , &&
\Sigma \simeq \frac{2}{\kappa_0^2} + \frac{\kappa_0^2}{\alpha_0^2} + \frac{2}{\alpha_0^2} \left(1-\sqrt{1+\kappa_0^2}\right) .
\end{align}
Assuming furthermore $\kappa_0 \ll 1$, we find that thin Q-shells sit at $\rho \sim\Sigma/(2\Omega_0^2)\sim\kappa_0^4/(8\alpha_0^2\Omega_0^2)$.
Since $\kappa_0 \leq 1/\Omega_0$, the Q-shells cannot become arbitrarily large but have a maximal radius for a fixed set of potential parameters. From Fig.~\ref{fig:Radii}, it is clear that this maximal radius is larger than the maximal radius of Q-\emph{balls}; upon integrating the profiles, we see that the maximal charge $Q$ of Q-shells also exceeds the maximal charge of Q-balls. Q-shells are thus configurations that hold more charge than a gauged Q-ball could.

\begin{figure}[t]
\includegraphics[width=0.49\textwidth]{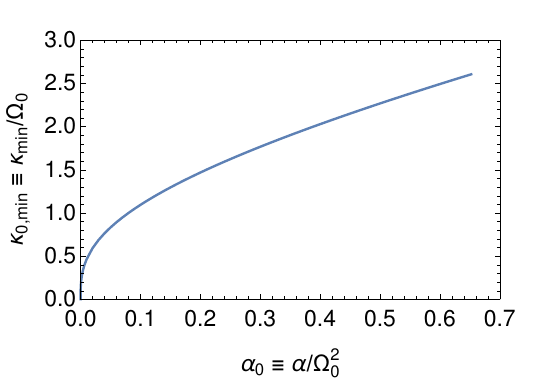}
\includegraphics[width=0.49\textwidth]{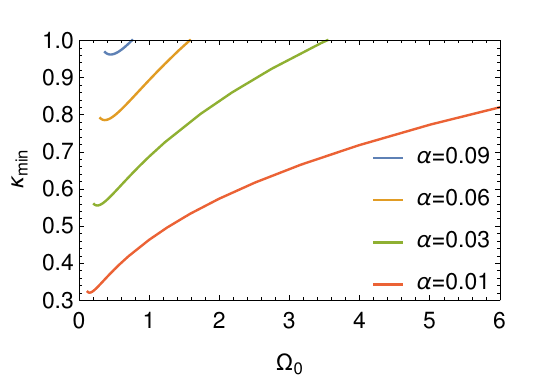}
\caption{
$\kappa_\text{min}$ is the smallest $\kappa$ for which Q-shell solitons exist according to the set of equations~\eqref{e.const1} and~\eqref{e.const2}.
Left: $\kappa_{0,\text{min}}$ as a function of  $\alpha_0$.
Right: $\kappa_\text{min}$ as a function of  $\Omega_0$ for some fixed $\alpha$.
}
\label{fig:kappamin}
\end{figure}

Unlike global Q-balls, gauged Q-ball solutions do not exist over the entire range $0<\kappa < 1$, as shown in Ref.~\cite{Heeck:2021zvk}, but instead have $\kappa_\text{min}\leq\kappa\leq1$.
The same is true for the Q-shell solutions, albeit with a different $\kappa_\text{min}$.\footnote{The Q-ball and Q-shell $\kappa_\text{min}$ merge for $\alpha_0\to 0$.} The dependence of the Q-shell $\kappa_\text{min}$ on the potential parameters can be extracted from Eqs.~\eqref{e.const1} and~\eqref{e.const2} and is shown in Fig.~\ref{fig:kappamin} (left).
For small $\alpha_0$, we find the excellent approximation
\begin{align}
 \kappa_{0,\text{min}} \simeq \left(9\alpha_0\right)^{1/3} \,,\label{e.kappamin}
\end{align} 
whereas $\kappa_{0,\text{min}} \simeq 3.21 \sqrt{\alpha_0}$ for larger values of $\alpha_0$.
Figure~\ref{fig:kappamin} (left) illustrates that Q-shell solutions only exist for $\alpha_0 < 0.65$, effectively putting a lower bound $\Omega_0 > 1.24 \sqrt{\alpha}$ for a given $\alpha$.
Similarly, we must  have $\kappa_\text{min} < 2.6\,\Omega_0$ to have any Q-shell solutions. 
Imposing the additional condition $\kappa_\text{min} \leq 1$ for localized soliton solutions gives an upper bound on the gauge coupling of the form $\alpha \leq 0.097$, see the right panel of Fig.~\ref{fig:kappamin}. 

The conditions $\alpha \leq 0.097$ and $\Omega_0 > 1.24 \sqrt{\alpha}$ restrict the allowed parameter space for Q-shells compared to Q-balls, which can exist for $\alpha$ up to $0.58$ and for $\Omega_0$ down to zero~\cite{Heeck:2021zvk}. 
In addition to the lower bound on $\kappa$ derived above for Q-shells and in Ref.~\cite{Heeck:2021zvk} for Q-balls, there exist upper limits on $\kappa$: for narrow Q-shells, this is simply $\kappa \leq 1$, or $\omega \leq m_\phi$; for wide Q-shells and thin-wall Q-balls, a more restrictive upper limit on $\kappa$ can be obtained in some regions of parameter space. As shown in Ref.~\cite{Heeck:2021zvk}, thin-wall Q-balls only exist when the maximum at $f_+(A(0))$ exists. The same constraint applies to wide Q-shells, leading to Eq.~\eqref{e.kappaMax}. 
For Q-balls the analysis is more complicated, but since both the wide Q-shell and the thin-wall Q-ball lie along the nearly the same path in the effective potential, we infer that when the wide Q-shell cannot exist neither can the thin-wall Q-ball.
 
Finally, a necessary condition for for physically stable solitons is $E < Q m_\phi$ in order to forbid the soliton decay into $Q$ scalars.
Between the Q-balls and Q-shells, we have up to four different solitons for the same set of potential parameters. This makes it possible, in principle, for some of the solitons to decay into more energetically favorable solitons. We do not attempt to study such instabilities here but leave this for future work.

\section{Results\label{s.results}}

We now compare our theoretical models with the exact numerical solutions. These solutions are obtained using finite element methods as outlined in Ref.~\cite{Heeck:2021zvk}. Seed functions for the numerical solutions are exactly the Q-shell ansatz outlined in the previous section. Both the ansatz and the exact solutions can be used to determine the charge $Q$ and energy $E$ of a given configuration using the Eqs. \eqref{e.charge} and \eqref{e.energy}, respectively. The radii of the numerical solutions are taken to be the point along a given transition where $f''(\rho)=0$, that is, $f''(R_{<,>})=0$.

\begin{figure}[t]
\includegraphics[width=0.48\textwidth]{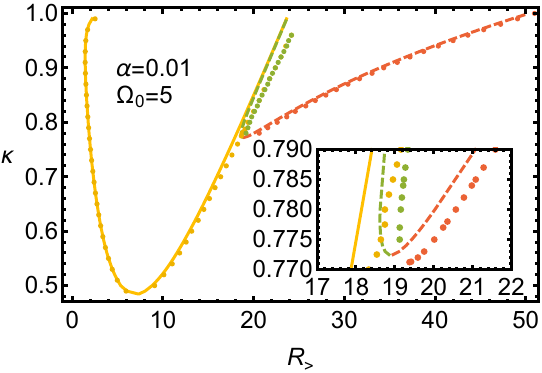}
\includegraphics[width=0.49\textwidth]{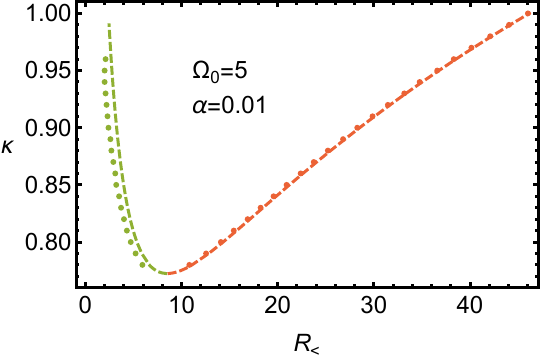}\\
\includegraphics[width=0.49\textwidth]{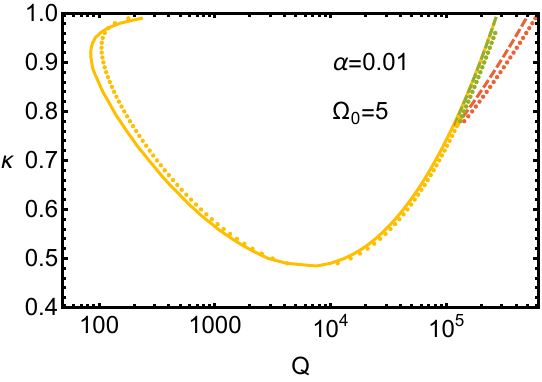}
\includegraphics[width=0.50\textwidth]{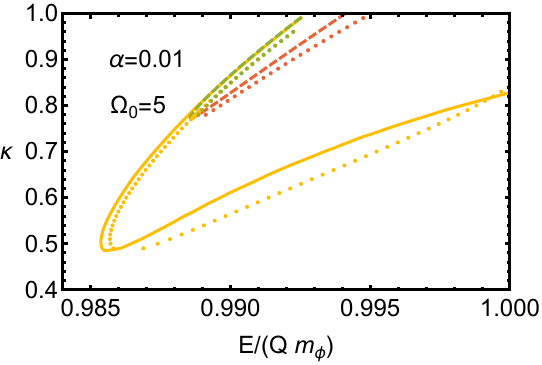}\\
\includegraphics[width=0.49\textwidth]{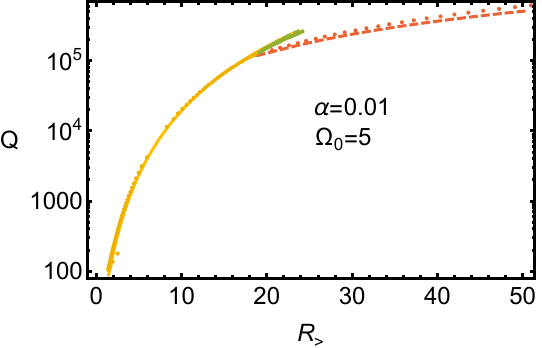}
\includegraphics[width=0.50\textwidth]{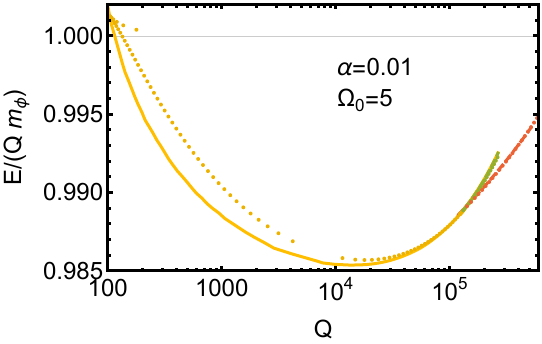}
\caption{
Theory prediction (solid lines for Q-balls~\cite{Heeck:2021zvk} and dashed lines for Q-shells) and numerical results (dots) for various gauged soliton characteristics. The wide (narrow) Q-shell results are shown in green (orange). The yellow lines and dots correspond to Q-balls.
}
\label{fig:Bmark1}
\end{figure}

Figure~\ref{fig:Bmark1} shows the results for Q-shells with $\Omega_0=5$ and $\alpha=0.01$ ($\alpha_0=4\times10^{-4}$). The wide (narrow) Q-shells are shown in green (orange). For reference, we also include the Q-ball solutions in yellow, obtained following Ref.~\cite{Heeck:2021zvk}. The theoretical Q-ball (Q-shell) predictions are given by the solid (dashed) lines, and the numerical solutions lie along the solid points. The general agreement between the theoretical predictions and the exact numerical solutions is excellent and is similar for other benchmark values as well.

Quantitatively, we find the Q-shell $\kappa_\text{min}\approx0.77$ from Fig.~\ref{fig:Bmark1} in agreement with our prediction from Eq.~\eqref{e.kappamin}.
For this parameter choice, we see that thin-wall Q-balls and wide Q-shells have the same $\kappa_\text{max}\approx0.97$. This is not captured by our theoretical prediction, as can be seen most easily in the plot of $R_<$ versus $\kappa$: the theory prediction for $R_<$ never reach below $9/4$, the constraint from Eq.~\eqref{e.kappaMax}, while the numerical results actually reach about $R_<\simeq 2$ before the solutions disappear. 
We see that in general the narrow Q-shells have larger radius ($R_>$), charge, and energy than the wide Q-shells and Q-balls of equal $\kappa$. The wide Q-shells lie along the regions of largest Q-ball radius, energy, and charge. The ratio of the energy to charge is quite similar for all of the solitons, so that narrow Q-shells are only mildly more stable than wide Q-shells and Q-balls of equal charge $Q$. 
Notice that Q-shells have $\partial Q/\partial \omega > 0$ just like thin Q-balls (Fig.~\ref{fig:Bmark1}), which, however, does not imply instability~\cite{Nugaev:2019vru}. A dedicated discussion of Q-shell stability is left for future work.

\section{Conclusion\label{s.con}}

Gauged solitons are classical field configurations of scalar fields that carry a $U(1)$ gauge charge. The best-known examples of such solitons are gauged Q-balls, which are straightforward generalizations of \emph{global} Q-balls. The dynamics of the gauge field makes possible qualitatively different solitons as well, taking the form of Q-\emph{shells}. Employing a generic sextic potential, we have shown here that the previously overlooked Q-shells exist over much of the parameter space and can carry higher charges, more energy, and larger radii than their Q-ball counterparts.
We have developed analytical approximations that describe these Q-shells remarkably well and also enable efficient numerical searches via finite-element methods.
Despite being studied for one particular scalar potential here, we expect Q-shells to arise in most, if not all, potentials that admit global Q-balls.
Q-shells are hence ubiquitous solitons, and studies of other potentials along the lines of our analysis should be useful and straightforward.

The Q-shells described here pass the most basic stability criterion, $E < m_\phi Q$, that forbids soliton decay into $Q$ individual scalars. Indeed, by this metric, the narrow Q-shells are slightly \emph{more} stable, for a fixed charge, than the Q-balls. A more detailed analysis of Q-shell stability is left for future work and must address both stability with respect to radial perturbations as well as possible decays of solitons with equal charge $Q$ into each other.

\acknowledgments
This work was supported in part by NSF Grant No.~PHY-1915005. C.~B.~V.~also acknowledges support from Simons 
Investigator Award \#376204. 

\bibliographystyle{utcaps_mod}
\bibliography{BIB}

\end{document}